# Study on the simulation control of neural network algorithm in thermally coupled distillation

ZhaoLan Zheng,Yu Qi

School of Chemical Engineering, East China University of Science and Technology, 200237 Shanghai, China.

## Abstract

Thermally coupled distillation is a new energy-saving method, but the traditional thermally coupled distillation simulation calculation process is complicated, and the optimization method based on the traditional simulation process is difficult to obtain a good feasible solution. The neural network algorithm has the advantages of fast learning and can approach nonlinear functions arbitrarily. For the problems in complex process control systems, neural network control does not require cumbersome control structures or precise mathematical models. When training the network, only the input and output samples it needs are given, so that the dynamics of the system can be controlled. Performance is approaching. This method can effectively solve the mathematical model of the thermally coupled distillation process, and quickly obtain the solution of the optimized variables and the objective function. This article summarizes the research progress of artificial neural network and the optimization control of thermally coupled distillation and the application of neural network in thermally coupled distillation.

**Keywords：Neural network algorithm, thermally coupled distillation, simulation**

# 1   Introduction

Artificial Neural Network Algorithm (ANN) is a simulation of the physical structure of the human brain, that is, the computer simulation method simulates the human brain from the physical structure so that the system has certain intelligence of the human brain. As we all know, the human brain is a complex biological network composed of billions of highly interconnected neurons. It is also the source of human analysis, association, memory, and logical reasoning capabilities. Neurons transmit information to each other through synaptic connections. The way and strength of the connection changes with learning, so that the learned knowledge is stored. The artificial neural network model composed of neurons that simulate the basic unit of information storage and processing in the human brain has intelligent behaviors such as self-learning and self-organization, which can make the machine have a certain level of intelligence. Neural network technology [1-2] has a strong ability to integrate information, can well solve the problem of complementarity and redundancy between input information, and can properly coordinate conflicting input information. Therefore, the neural network technology has shown its superiority in the design of multi-variable, large-scale and complex control schemes. It has been widely used in many fields such as intelligent driving, signal processing, drug screening, wind power prediction, etc., especially in chemical process control, and it is one of the first fields to apply neural networks. Antsaklis [3] et al. loaded the physical quantities from various measuring instruments (such as temperature, acidity, concentration, viscosity, vaporization speed, products, etc.) from various measuring instruments into the trained neural network during the chemical reaction process, The network can predict abnormal situations in time to stop the occurrence of abnormal situations in time. The good adaptability and self-learning performance of artificial neural network in processing high-dimensional, non-linear process industry data can provide petrochemical enterprises with production operating system optimization modeling, device optimization, dynamic and static equipment fault diagnosis, and energy saving, safety, environmental protection and pre-efficient support [4-9].

In the petrochemical process, the rectification process is one of the most widely used unit operations, and also the unit operation with the largest energy consumption. Improving the energy utilization rate of the rectification process is always a hot research topic. Thermally coupled distillation [10] has attracted widespread attention because of its characteristics of thermal coupling and equipment integration, which can save energy and equipment investment. Thermally coupled rectification changes the composition distribution in the tower through the coupling of materials and solves the problem of backmixing of intermediate components. It is an internal thermal coupling enhancement technology [11], which can improve thermodynamic efficiency, reduce process energy consumption, and reduce the entire process Production costs. The traditional method of simulating and optimizing the thermally coupled rectification process is to decompose the system first, and then perform the convergence calculation of the main and auxiliary towers in turn. It is difficult to achieve the overall matching and optimization of the main and auxiliary towers in the calculation process. Therefore, improving the simulation and optimization of the thermocouple process has important theoretical and practical significance. The neural network model can simulate this convergence calculation well. This paper briefly describes the basic model of neural network and the optimization control research of thermally coupled distillation and summarizes the application of neural network in thermally coupled distillation.

## 2 The principle and application development of neural network

Artificial neural network is based on the basic principles of neural networks in biology. After understanding and abstracting the human brain structure and external stimulus response mechanism, it uses network topology knowledge as the theoretical basis to simulate the complex information processing mechanism of the human brain's nervous system. A mathematical model. Generally, an artificial neural network is composed of three basic elements: neurons, topological structure and learning rules. Neurons form a basic network-like topology. After the information is input through the neurons, it is processed according to certain learning rules, and the results are output through the neurons [1]. Compared with traditional computing, artificial neural networks have significant parallel processing, sub-storage, convergence, associative memory and other functional characteristics. Artificial neural networks can be divided into continuous and discrete types according to their functions. According to the topology, they can be divided into forward networks and feedback networks. According to learning rules, they can be divided into supervised learning networks and unsupervised learning networks. Commonly used neural network models are mainly BP, RBF, Hopfield etc.

The development process of neural network can be roughly divided into two periods before 2006[12]. In the 1940s, Warren McCulloch and Walter Pitts opened the door to research in the field of artificial neural networks. They proved in principle that artificial neural networks can calculate any arithmetic and logical function [12]. In the late 1950s, although many people were influenced by Minsky and Papert, there was no powerful computer to support various experiments and the perceptron was questioned, causing the research of artificial neural networks to stagnate for more than ten years [12]. The 1980s became a turning point in the development of neural networks. Nobel Prize winner Hopfield John proposed the Hopfield neural network model [12]. The dynamic nature of this Recurrent neural network may be used to solve complex problems. In 1974, Werbos [13-14] proposed the BP (Back Propagation) algorithm for neural network learning in his doctoral thesis, which provided a practical solution for the learning, training and implementation of multilayer neural networks. At the same time, in 1986, a team of scientists led by Rumelhart and McCelland [15-16] conducted a detailed analysis of the error back propagation algorithm of the multilayer network, which further promoted the development of the BP algorithm. The topological structure of BP network includes input layer, hidden layer and output layer. It can store this complex mapping relationship through learning without knowing the specific mathematical expressions of the input and output in advance. The learning of the parameters in the network usually adopts the strategy of back propagation and uses the fastest gradient information to find the smallest network error. The combination of parameters. BP algorithm is currently the most popular fault diagnosis feedforward network [17].

Following BP, in order to simulate the local response characteristics of biological neurons, Broomhead and Lowe [18] introduced radial basis functions into the design of neural networks in 1988, forming a radial basis function neural network RBF. Later, Jackson and Park [19-20] demonstrated the uniform approximation performance of RBF on nonlinear continuous functions in 1989 and 1991, respectively. The RBF neural network is a three-layer forward network. Its basic working principle is: use the hidden layer space formed by RBF to project the low-dimensional input vector and transform the data into the high-dimensional space so that the original linearity is inseparable The problem can become linearly separable. Since the input layer only plays the role of signal transmission in the RBF network, the connection weight between the input layer and the hidden layer is 1, the hidden layer realizes the nonlinear projection of the input features, and the output layer is responsible for the final The linear weighted summation. The RBF network has a fast learning convergence rate. One important reason is that it belongs to a local approximation network and does not need to learn the weights of hidden layers, which avoids the time-consuming process of layer-by-layer transmission of errors in the network. The RBF network is also an important sign of the real practical application of neural networks. It has been successfully applied to engineering fields such as nonlinear function approximation, pattern classification, control system modeling, time-varying data analysis, and fault analysis and diagnosis [20-23].

In the past 10 years, artificial neural networks have achieved rapid development. In 2006, Professor Hinton [24] provided an effective algorithm for initializing weights, which led to the development of deep learning in the field of research and applications [25-27].

## 3    Application development of thermally coupled distillation

Thermally coupled distillation is one of the most commonly used distillation enhancement technologies and one of the energy-saving distillation processes. Thermally coupled rectification changes the composition distribution in the tower through the coupling of materials and solves the problem of backmixing of intermediate components. It is an internal thermal coupling strengthening technology, which can improve thermodynamic efficiency, reduce process energy requirements, and reduce the production cost of the entire process . It is mainly used to separate three-component mixtures or divide the mixture into three products. It can be divided into the following forms: ①Sideline distillation tower, composed of main tower and sideline rectification tower; ②Sideline stripping tower, composed of main Tower and side-line stripping tower; ③Completely thermally coupled distillation tower, this tower was first proposed by Petlyuk, so it is also called Petlyuk distillation tower, composed of main tower and pre-fractionation tower, the role of pre-fractionation tower is to carry out the mixture After preliminary separation, the light key components are all separated from the top of the tower, the heavy key components are completely extracted from the tower kettle, and the intermediate components are distributed between the top and bottom of the tower. The main tower serves to pre-separate the top of the tower. It is further separated from the material at the bottom of the tower to obtain the product that meets the requirements; ④Vertical partition rectification tower, the vertical partition is used inside the tower to divide the tower into two parts. This structure can be considered essentially Combine the main tower and pre-splitting tower of the petlyuk rectification tower in the same tower. For a given material, the partition tower rectification requires a smaller reflux ratio than the conventional rectification process, which increases the operating capacity, saves energy by up to 60%, and saves equipment investment by 30%.

People have carried out a lot of design, optimization and application research for the above various forms of thermally coupled distillation. Waybum and Seader [28] proposed a homotopy consensus algorithm suitable for difficult convergence problems. The main purpose of the homotopy continuation method is to address the shortcomings of Newton's method for the initial value of the iteration. It can guarantee Large-scale convergence of the initial value. Aiyue Zhou [29] and others used the homotopy consensus algorithm to simulate the thermally coupled distillation process. Through the simulation calculation of the thermally coupled distillation column for separating benzene, toluene, and o-xylene with multiple solution conditions and multiple working conditions, When the linear system generated by Newton's method and homotopy continuation method, an improved THOMAS method and progressive block elimination method (BRR) combined algorithm is proposed.Xien Xu [30] used the block tridiagonal matrix method of Naphtali-Sandholm to complete the simulation of the rectification process in which the liquid phase may be stratified on part or all of the equilibrium stages in the thermocouple column system. Ping Qu [31] took the butadiene separation device as the object and conducted a simulation analysis and research on the operating characteristics and energy-saving effects of fully thermally coupled distillation for non-ideal systems. Deming Yang [32] used reflux ratio and coupled vapor-phase stream flow as tuning parameters to determine the best design conditions for the minimum energy consumption and stable operation of the all-thermal coupling tower to separate paraffins. Hong Li [33] and others developed a new type of differential pressure thermally coupled rectification technology to solve the problem of incomplete thermal coupling between the top condenser and the bottom reboiler.

The most well-known TCDS sequence, the Petlyuk configuration, has some operational challenges due to bidirectional vapour flow, which makes its implementation difficult in two-column mode. To overcome these limitations, a number of unidirectional vapour flow configurations have been proposed in the literature [34].

Mansour Emtir, Endre Rev & Zsolt Fony [35] studied the energy-saving properties of all column sequences in three-component rectification separation through rigorous simulation. Fidkowski & Agrawall [36] focused on the energy consumption characteristics of the four-component thermal coupling distillation separation under the minimum reflux ratio, and still came to the conclusion that this distillation structure is the best, and the energy saving reached 20.50%. In the simple design method proposed by Amminudin et al [37-38], more rigorous vapor and liquid balance models are emphasized; Kim [39-40], starting from the degree of freedom of the analysis system, proposed a rigorous method for designing a completely thermally coupled tower; Muralikrishna et al [41] proposed a diagram method to design the dividing wall tower. Jiangwei Xie [42] et al. proposed a method to use response surface methodology (RSM) instead of non-dominant genetic algorithm (NSGA-II) for multi-objective optimization design of the next-door tower.

Based on the stochastic optimization strategy [43-44], Weizhong An [45-46] proposed a decomposition solution strategy for the uncertainty of the number of columns required by the thermally coupled complex distillation process system and the number of condensers and reboilers. The original problem is decomposed into a series of sub-problems with different numbers of tower sections to be solved separately.

For the optimization control problem of high-purity thermally coupled distillation column, Yingxiao Zhang [47] proposed a control scheme based on nonlinear process model control (NPMC) for a nonlinear object model-benzene-toluene system model, and The conventional PID control schemes are compared, and the results show that NPMC is one of the most effective control schemes [48]. It adopts the principle of general model control (GMC) [49] to design the nonlinear process basic model controller, which can smoothly realize the change of the set value and effectively eliminate the external interference.

A direct advantage of using thermal coupling technology is to reduce the number of heat exchangers in the system, which is more attractive for separating multi-component mixtures. Petlyuk [50] extended the idea of three-component total thermal coupling distillation to separate N-component mixtures, and proposed that no matter the number of separated components, the entire system can only use one condenser and one reboiler. Triantafyllou and Smith [51] optimized the flow value of each part of the vapor and liquid phases under a given number of plates, and only took the recovery rate as the adjustment variable. Hernandez and Jimenez [52] proposed to use dynamic model to optimize the design of coupling tower.

# 4    Application of Neural Network Algorithm in Thermally Coupled Distillation

At present, intelligent optimization algorithms are used more in the online optimization of chemical processes. In the research of artificial neural networks, there are two kinds of stochastic optimization algorithms that are very popular. They are the Simulated Annealing Algorithm (Simulated Annealing, SA) proposed by Metropolis [53] and the Genetic Algorithm (GA) proposed by Holland [54]. SA simulates the annealing process of metal materials after heating, and GA simulates the natural evolution process of organisms. From the perspective of simulation objects, the two have almost nothing in common, but from the formal structure of the algorithm itself, the two are extremely similar. SA ensures the probabilistic reachability of each point in the state space by means of the Generation Function, simulates the equilibrium state by the Acceptation Operator, and guarantees the directionality of the algorithm iteration process by accepting the directional changes of the operator; while GA The synergy of crossover operator and mutation operator is used to ensure the probability reachability of each point in the state space, and the direction of the algorithm iteration process is ensured by the action of selection operator. From a mathematical point of view, random algorithms for global optimization can be divided into two categories: one is through traversal search, such as SA and GA, etc.; the other is through directional advancement, such as generalized genetic algorithms [55].

## 4.1    BP neural network based on genetic algorithm

Since thermally coupled rectification uses a complex series of main towers and auxiliary towers to replace conventional distillation towers, if traditional methods are used to simulate a large number of iterative calculations between the main and auxiliary towers, the calculation process is complicated and the established model lacks simplicity It is not conducive to further optimization calculation of the process. Yanmin Wang and Li He [56-57] used artificial neural networks and genetic algorithms to propose a new modeling method and optimization algorithm for the thermally coupled distillation separation process, and used artificial neural networks to establish a thermally coupled distillation process model. The genetic algorithm optimizes the thermally coupled distillation process. They used actual engineering examples to study the separation of carbon five components and the separation of butadiene respectively. The results showed that this method can not only effectively and conveniently solve the mathematical model of the thermally coupled distillation process, but also quickly obtain Optimize variables and objective functions, and have the ability to obtain global optimal solutions.

On the basis of Aspen simulation software, ANN is used to establish a thermally coupled distillation process model. The steps are: (1) Confirmation of input and output variables. The system process analysis determines the target variables and constraints and the optimization variables of GA, and then determines the input and output variables of ANN. (2) Acquisition of ANN training data. Using Aspen PLus simulation calculation, several sets of data with a certain interval and range are obtained. (3) Training of ANN weights. Given network structure, convergence accuracy E... Use part of the data calculated in (2) to train the ANN. (4) ANN accuracy test. Use the remaining data in (2) to detect the training effect of the artificial neural network. Using BP algorithm, selecting enough sample data and reasonable network environment for training, can meet the simulation requirements. However, the BP algorithm is a gradient-based method. This method has slow convergence speed and is often troubled by local minima. GA absorbs the biological evolution principle of "survival of the fittest, survival of the fittest", and can be used without specific information about the object Perform global optimization in a complex space. The basic principle of the application of artificial neural network based on genetic algorithm [58] is to use genetic algorithm (GA) to optimize the connection weights of neural networks, and use GA's optimization ability to obtain the best weights. Because genetic algorithms have the advantages of strong robustness, randomness, globality and suitable for parallel processing, they are widely used in neural networks, and there are many successful applications. The black box mathematical model is first established by artificial neural network, and then the genetic algorithm is used for global optimization, which solves the complex calculation process of traditional thermally coupled distillation simulation and it is difficult to optimize the overall optimization between the main and auxiliary towers.

## 4.2   RBF neural network based on hybrid optimization algorithm

If the center of the radial basis function (RBF) neural network is selected properly, the hidden layer only needs a few neurons to get a good approximation effect, and the learning speed is fast, and it also has the characteristics of approaching the optimal point. Kun Cai and Xinggao Liu [59] proposed a hybrid optimization algorithm based on RBF neural network internal model control, which is a hybrid optimization algorithm of particle swarm optimization. Through actual engineering examples, the benzene-toluene system is used as an example to study. Through comparison, the optimization method can overcome the shortcomings of large PID overshoot and long response time, and it is smaller than conventional RBF-IMC The overmodulation also overcomes the shortcomings of IECR-IMC [60] poor stability. The RBF neural network internal model control scheme based on the hybrid optimization algorithm is an excellent internal thermally coupled distillation control strategy.

### 4.2.1 Radial basis function neural network dynamic system modeling and realization of control strategy

According to the composition principle of the internal model controller, the positive model (internal model) and the inverse model (controller model) of the controlled object are first established. The radial basis function (RBF) neural network can naturally divide the input space, so that the input vector can be extended to the high-dimensional hidden unit space, so that the local minimum is greatly reduced. Through pre-sampling the input and output data of the system, establish NNlVl and NNC models offline through RBF neural network, and then put the learned network into the actual control system to run; the system should continuously check the input and output data during actual operation, according to the positive model The weight of NNM is corrected online with the object output error, and the weight of NNC is corrected online according to the set value and object output error, and the parameters of the two networks are fine-tuned. This not only enables the system to have a faster learning speed, but also When the given reference input changes, the system can still track the given reference input well, thus realizing the adaptive control of the controlled object.

### 4.2.2 Improved algorithm of radial basis function (RBF) neural network

The subtractive clustering algorithm [61] is a simple and effective clustering algorithm. Compared with other methods, this method does not need to determine the number of clusters in advance. K-means [62] algorithm is a basic division method in clustering algorithms. The subtractive clustering algorithm regards each data sample as a potential cluster center, and determines the cluster center according to the density index of the sample data, which can effectively reflect the distribution of the data. The subtractive clustering algorithm can determine the number of class centers, but it can only select the class centers within the sample range, but cannot modify them; the K-means algorithm can continuously modify the class centers, but the number of clusters K is a given by experience The K-means clustering result is greatly affected by the initial value, and the function center point generated by the K-means clustering algorithm, as well as the randomly generated RBF neural network weight and Gaussian function radius r. The PSO algorithm and its extension mean-filed filtering algorithms [63-64] are based on swarm intelligence, which has good optimization capabilities, but if the initial particles deviate too much from the target population, it will require a larger iteration time cost. Therefore, the clustering number K of the K-means algorithm can be determined by the subtractive clustering algorithm, and then the multiple clustering of the K-means algorithm and the optimization of the gradient descent method can produce a more reasonable improved algorithm for the initial particle swarm of the PSO algorithm.

## 5   Conclusion

The difficult control problems in the production process of the rectification tower are correlation, complexity and uncertainty. The traditional control method cannot meet the requirements of production control. The superiority of neural network in the field of control lies in: neural network can approximate arbitrary nonlinear mapping with arbitrary precision, and brings a new and non-traditional expression tool to the modeling of complex systems; multi-input and multi-output structure The model can be easily applied to multivariable control systems; at the same time, because it can integrate qualitative and quantitative data, it can use the structure of connectionism, combined with traditional control methods and symbolic artificial intelligence. It can be predicted that in the near future, the neural network algorithm will be more widely used in the optimization of thermally coupled distillation control.